\documentclass[journal]{vgtc}                     %

\newcommand{\fix}[1]{\textcolor{black}{{{#1}}}}
\newcommand{\add}[1]{\textcolor{black}{#1}}

\onlineid{1566}

\vgtccategory{Research}

\vgtcpapertype{Theoretical \& Empirical}

\title{
Characterizing Visualization Perception with \fix{Psychological Phenomena}: Uncovering the Role of \textit{Subitizing} in Data Visualization
}

\author{%
  Arran Zeyu Wang,
  Ghulam Jilani Quadri, Mengyuan Zhu, Chin Tseng, and 
  Danielle Albers Szafir
}

\authorfooter{
  \vspace{-1mm}
  \item
  	Arran Zeyu Wang, Mengyuan Zhu, Chin Tseng, and Danielle Albers Szafir are with the University of North Carolina at Chapel Hill. E-mail: \{zeyuwang, chint, gisellez, danielle.szafir\}@cs.unc.edu
  \item
  	Ghulam Jilani Quadri is with the University of Oklahoma and the University of North Carolina at Chapel Hill.
  	E-mail: quadri@ou.edu
}

\shortauthortitle{Subitizing}

\teaser{
\centering
\includegraphics[width=0.95\columnwidth]{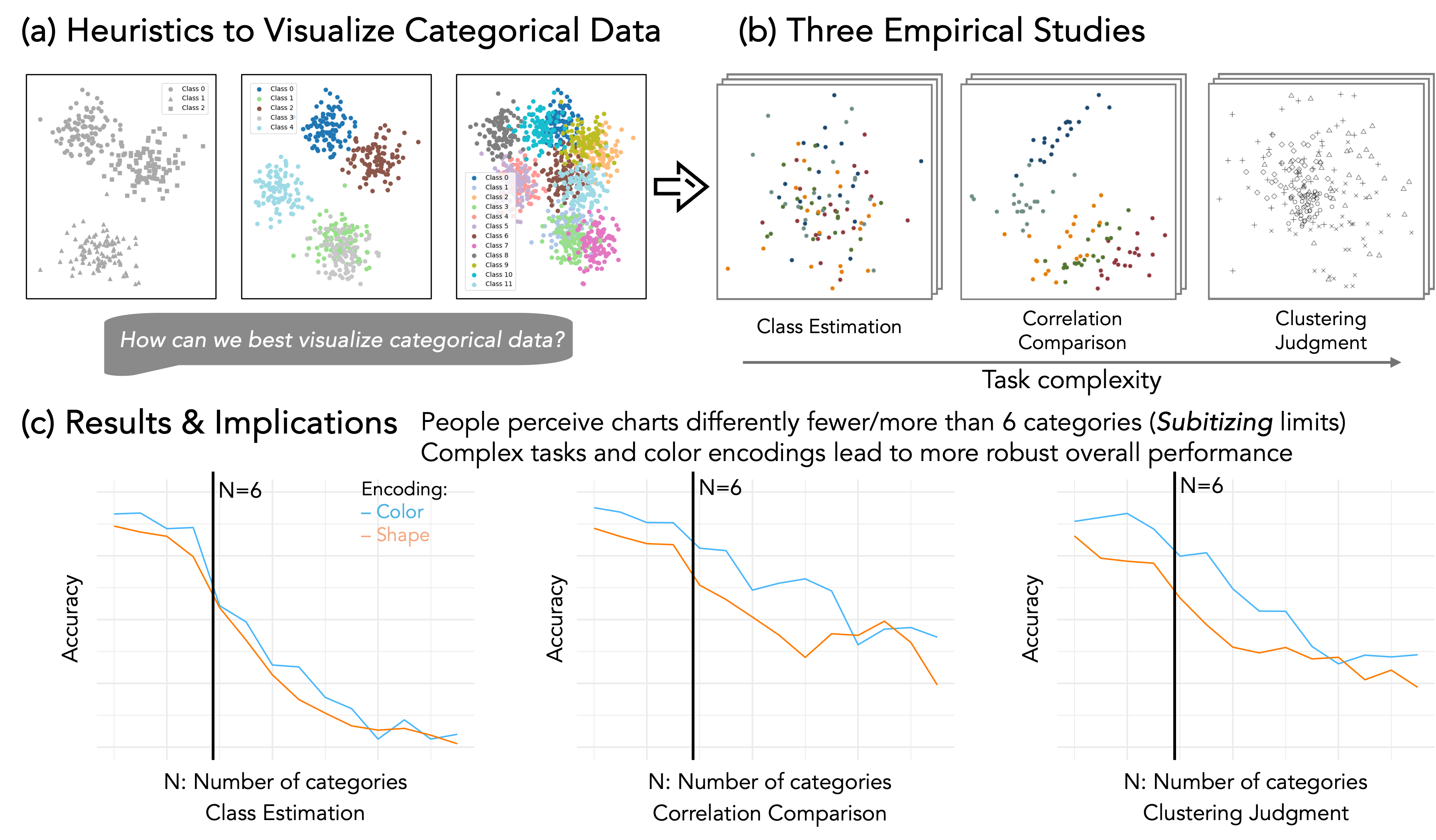}
\vspace{-0.5em}
\caption{ 
\add{
We explored the potential of using psychological theories to validate visualization design heuristics by assessing a psychological
phenomena called \emph{subitizing} in 
categorical data visualization.
(a) There are many existing heuristics when visualizing categorical data, varying from limiting designs to 5 to 12 categories.
(b) We designed three empirical studies to characterize performance from 2--15 categories with color and shape encodings in visualizations with three experimental tasks: class estimation, correlation comparison, and clustering judgments.
(c) Our results provide empirical evidence that people perceive charts differently before and after 6 categories with significant performance reduction. 
We term this threshold as the \emph{Subitizing} limit, meaning the subitizing phenomenon can help people immediately perceive information from charts with fewer than 6 categories.
More complex aggregation-based tasks and alternative encodings influence the effect of subitizing.
}
}
\label{fig:teaser}
}
\abstract{
\add{Understanding how people perceive visualizations is crucial for designing effective visual data representations; however, 
many heuristic design guidelines are derived from specific tasks or visualization types, without considering the constraints or conditions under which those guidelines hold.
In this work, we aimed to assess existing design heuristics for categorical visualization using well-established psychological knowledge.}
Specifically, we examine the impact of the \emph{subitizing} phenomenon in cognitive psychology---people's ability to automatically recognize a small set of objects instantly without counting---in data visualizations.
We conducted three experiments with multi-class scatterplots---\add{between 2 and 15 classes with varying design choices}---across three different tasks---class estimation, correlation comparison, and clustering judgments---to understand how performance changes as the number of classes (and therefore set size) increases.
Our results indicate if the category number is smaller than six, people tend to perform well at all tasks, providing empirical evidence of \emph{subitizing} in visualization. When category numbers increased, performance fell, with the magnitude of the performance change depending on
task and encoding.
\add{Our study bridges the gap between heuristic guidelines and empirical evidence by applying well-established psychological
theories, suggesting future opportunities for using psychological theories and constructs to characterize visualization perception.}
} %


\graphicspath{{figs/}{figures/}{pictures/}{images/}{./}} %

\usepackage{tabu}                      %
\usepackage{booktabs}                  %
\usepackage{lipsum}                    %
\usepackage{mwe}                       %

\usepackage{mathptmx}                  %
\usepackage{amsmath,amsfonts}
\usepackage{algorithmic}
\usepackage{algorithm}
\usepackage{array}
\usepackage{textcomp}
\usepackage{stfloats}
\usepackage{url}
\usepackage{verbatim}
\usepackage{graphicx}
\usepackage{cite}  
\usepackage{amsmath}
\usepackage{wrapfig}

\begin{document}

\maketitle

\section{Introduction}
\label{sec-intro}

\fix{In visual data communication}, effective visualizations enable users to interpret complex information quickly and accurately~\cite{franconeri2021science}.
As the size and complexity of a dataset grows, 
designing visualizations that are both efficient and accurate becomes increasingly important, but also increasingly challenging~\cite{szafir2023visualization, szafir2018good, franconeri2021science}.
For example, most complex datasets include both numeric and categorical data. While past studies have extensively investigated methods for representing traditional numeric data~\cite{saket2018task, behrisch2018quality}, guidelines for categorical data remain largely heuristic and relatively
sparse~\cite{munzner2014visualization, tseng2023evaluating}.

One heuristic for categorical data is to encode no more than six categories, as stated in Adobe Design Guidelines~\cite {adobedesign} and elsewhere without supporting evidence or a clear origin \cite{graze2024building}. 
When the category numbers in a dataset increase, the complexity of the visualization increases as well, often leading to a significant descent of the \fix{\emph{perception effectiveness}}\footnote{\fix{In the context of this study, ``perception effectiveness'' is operationalized as the accuracy with which people can extract specific information about categorical data presented in multiclass scatterplots within a given time constraint.}} in visualizations~\cite{haroz2012capacity, tseng2023evaluating}.
However, modern datasets frequently contain far more than six categories, forcing designers to either reduce the granularity of categorical data by grouping related categories~\cite{kogan2006grouping, alsallakh2016state}, repeat encodings or facets over multiple categories~\cite{deng2022revisiting, heer2009sizing}, or choose categorical encoding designs that violate this heuristic like extended color ramps or palettes~\cite{smart2019color, yuan2020evaluation}.
More recent studies suggest that, for certain tasks, people are still able to reliably reason over more than six distinctly encoded categories~\cite{tseng2023evaluating}; however, these studies also reveal a steep performance decline near six categories.
In this paper, we explore the six-category heuristic over a range of visualization tasks to determine its generalizability and infer potential perceptual grounding for this common categorical design guideline.

The decline in performance after six categories is not just a matter of aesthetics~\cite{palmer2013visual}; it directly impacts the user's ability to make accurate and efficient judgments based on the visualized data~\cite{tseng2023evaluating}.
The heuristic itself is not always employed consistently, with past work recommending a wide range of thresholds, including six~\cite{adobedesign}, five to seven~\cite{peterson2013gestalt}, eight~\cite{wong2010points}, six to nine~\cite{intelligaia}, ten~\cite{tenfusion}, eleven~\cite{boynton1989eleven}, or six to twelve categories~\cite{munzner2014visualization}.
One likely source of variance in these recommendations is the lack of theoretical or empirical basis. 
We approach this heuristic through the lens of psychological theories and constructs.
In cognitive psychology, the phenomena of \emph{subitizing} allows people 
to instantly and accurately recognize and quantify small numbers of objects---typically around four to six---without counting~\cite{kaufman1949discrimination}.
For more than six objects, individuals must rely on counting or estimation, leading 
to decreased accuracy and 
increased cognitive load.
Past work has speculated that subitizing may play a role in categorical data interpretation in visualizations,
posing challenges to people's perception effectiveness~\cite{tseng2023evaluating, haroz2012capacity}. However, performance for categorical encodings tends to decrease non-linearly with the number of categories, potentially influencing the range of prescribed categories in past heuristics. 
These variations may be explained by another psychological theory, \emph{Fechner's Law}, 
which states that as the \emph{intensity} of a stimulus \fix{(in the context of visualization, the ``\emph{intensity}'' 
corresponds to the amount of information, such as point and category number, being displayed, see \autoref{fig:f-law})} increases, the corresponding perceived difference increases logarithmically~\cite{fechner1948elements}. 
\fix{
We specifically focused on the finding that, as the number of categories grows, the perceptual difference between them diminishes logarithmically, making it harder for people to distinguish between categories~\cite{goldstone2010categorical}.}
This may mean that the most dramatic performance decreases happen around the subitizing threshold and level off to stable perceived differences above a secondary threshold.

In this work, we investigated the influence of subitizing in data visualization by empirically examining 
how increasing the number of categories affects different data analysis tasks. 
We conducted a series of three crowdsourced experiments that tested 
people's abilities to \fix{comprehend visual information} across categories using multi-class scatterplots with varying numbers of categories (2 to 15), categorical encodings (color and shape), and visual complexities (point numbers) on three tasks: class estimation, correlation comparison, and clustering judgments (see \autoref{fig:teaser} (b)).
Our results indicated that when scatterplots encode fewer than six categories, 
accuracy remains stable and relatively high across all tasks.
This finding aligns with the concept of subitizing, suggesting that the human visual system is highly efficient at processing small sets of categories.
As the number of categories increases beyond this threshold, perceptual accuracy first significantly declines and then converges to a more stable performance, aligned with Fechner's Law.
However, aggregation-based tasks and color encodings remained more robust to increasing category numbers, suggesting additional visual mechanisms may be at play when engaged in these tasks. 
\autoref{fig:teaser} (c) summarizes the main implications.
By grounding our results in established subitizing phenomena in cognitive science, we not only contribute to the field of data visualization but also demonstrate the potential for leveraging cognitive psychology to inform and improve visualization design.
This grounding enables us to use our results to propose more generalizable and empirically-grounded guidelines that can help 
create more effective visualizations for categorical data.

The main contributions of this paper include:

\begin{itemize}
    \item \textbf{Empirical investigation of subitizing limits in visualization and design guidelines.}
    We provide empirical evidence that the perceptual accuracy in visualizations significantly declines when the number of categories reaches six, corresponding to  
    the psychological phenomenon of subitizing, and allowing for more generalizable, empirically-grounded guidance on categorical encoding based on the number of categories present in a dataset. 
    
    \item \textbf{Characterizing performance patterns 
    for categorical encodings within and beyond subitizing limits.}
    We find that performance remains relatively stable for fewer than six categories, consistent with subitizing, and decreases asymptotically for more than six categories, consistent with  
    people requiring more complex cognitive processes to comprehend the visualized data
    and matching Fechner's Law.

    \item \textbf{\fix{Examining visualization perception through the lens of psychological theories and constructs.}}
    \add{Our results, as a whole, further show how widely established \fix{psychological theories and constructs} can be used to measure the capabilities and performance for \fix{perception and cognition} in data visualization \fix{from a continuum of low-level, relatively invariant to high-level, relatively variant psychological phenomena} and ground the generalizability of visualization guidelines.}

\end{itemize}

\add{
Providing theoretical support and empirically grounded evidence for established heuristics can be crucial for visualization research and practice~\cite{lee2019broadening}.
Rather than establish new guidance, we aim to empirically deconstruct the underlying graphical perception process behind existing heuristics in categorical data visualization to better understand their application in practice and provide a research template for examining other design heuristics, responding to the call for better understanding the mechanisms behind how people accomplish different tasks~\cite{kosara2016empire}.
}

\section{Background}
\label{sec-related}

Research on categorical perception in  visualization emphasizes the importance of visual encodings like color and shape for effective data interpretation.
These studies are often tightly integrated with studies in psychology, particularly from cognitive and vision science, that highlight people's ability to recognize and perceive visual quantities.
Our work builds on these foundations and provides new insights into design choices for visualizing categorical data by empirically testing how robustly people conduct 
high-level tasks in multiclass scatterplots. 

\subsection{Scatterplots for Data Communication}

Scatterplots are an intuitive and widely-used visualization for bivariate quantitative data~\cite{sarikaya2018scatterplots, franconeri2021science}.
They are also one of the most studied visualizations, with experiments showing how different scatterplot designs support 
assessing trends~\cite{correll2017regression}, correlation perception~\cite{harrison2014ranking}, causal inference~\cite{wang2024empirical}, clustering~\cite{jeon2023clams}, and detecting outliers~\cite{sarikaya2018design}.  
Their visual simplicity and ability to support a range of tasks make them a common stimulus for visualization research to provide insights into general paradigms and implications of visualization design, similar to a ``fruit fly'' in biology~\cite{rensink2018information}.

Recent studies have emphasized scatterplots' effectiveness in communicating categorical information \cite{tseng2023evaluating, gleicher2013perception}. For example, scatterplot designs are often used to analyze class structures in dimensionally-reduced data~\cite{van2008visualizing}.
Categorical data visualization is a fundamental aspect of data representation, playing a crucial role in how users interpret and make decisions based on complex information~\cite{munzner2014visualization}. 
Representing categories differs from traditional numeric encoding as people must often select for and then compare patterns across different categories.
As the number of categories grows, people may struggle to distinguish between them due to factors like reduced discriminability and increased clutter, leading to errors in interpretation and perceptual judgments~\cite{haroz2012capacity, tseng2023evaluating}.

Scatterplots typically delineate categorical data using mark encodings, such as shape~\cite{burlinson2017open}, size~\cite{hong2021weighted}, texture~\cite{mayorga2013splatterplots}, or color~\cite{schloss2018color}. Encoding choice 
can significantly impact the ease with which people can interpret and analyze data.
Color and shape are the most commonly used encodings for categorical data~\cite{ware2012information, franconeri2021science}.

Color is a common encoding choice for categorical data due to its intuitive 
perceptual qualities: color provides a robust perceptual cue for grouping related objects \cite{schulz2003time}.
Hue, lightness, and perceptual distances have a significant effect on the efficiency of categorical color palettes in categorical visualizations~\cite{tseng2023evaluating,gramazio2016colorgorical}.
Color schemes specifically designed for categorical data significantly outperform sequential and diverging palettes, further emphasizing the importance of using distinct and easily distinguishable colors~\cite{tseng2024revisiting}.
However, as the number of categories increases, the perceptual distinctiveness of colors diminishes, leading to reduced accuracy~\cite{tseng2023evaluating, tseng2024revisiting}. Common heuristics for color encoding recommend using no more than six colors to represent data~\cite{adobedesign}. However, as with many common design heuristics \cite{kosara2016empire}, we lack empirical and mechanistic insight into these approaches to inform visualization design and to understand when, where, and how these guidelines and approaches should be applied. For example, alternative design strategies often leverage manipulations of lightness or saturation to successfully extend these palettes beyond six categories~\cite{graze2024building}.

Shape is another widely-used encoding for categorical data, offering a complementary approach to color.
Empirical evidence shows that the choice of shape palette also affects encoding effectiveness. For example, closed shapes may be more effective when representing two-class scatterplots compared to open shapes or mixed closed-open shape pairs~\cite{burlinson2017open}.
Shape can be particularly useful when combined with other encodings, such as color or size, to represent multiple dimensions of data simultaneously~\cite{smart2019measuring}.
However, 
Tseng et al.~\cite{tseng2024shape} found that when visualizing complex categorical data, 
shape encodings may be less effective overall than color and that shape's effectiveness may also be limited by the number of distinct shapes that can be easily recognized and differentiated by people~\cite{tseng2024shape}. Unlike color, we lack known heuristics for shape palette design, especially with respect to the number of potential categories.

For both color and shape, past studies show that patterns in effectiveness vary non-linearly as a function of the number of categories; however, these studies focus on averaging tasks and provide limited insight into the broader implications of this performance. 
This study investigates patterns in categorical data analysis in scatterplots across a wider range of tasks. Our objective is to develop more robust design guidelines and formulate theoretical hypotheses about task performance, thereby informing more effective design strategies.

\subsection{Psychology in Visualization Research}

Data visualization has increasingly drawn on principles from psychology to inform design practices~\cite{elliott2020design, szafir2023visualization, bae2025bridging}.
Many \fix{theories, phenomena, and constructs} in psychological research provide valuable insights into how people perceive and process visualized quantities that can guide more effective design practice and ground past heuristics through translational research \cite{szafir2016four}.
These laws describe the relationship between the physical properties of stimuli (i.e., a visualization) and 
corresponding percept (i.e., an inference about the data), offering a framework for understanding the limits of human perception that, when applied to visualization, offers novel means for grounding and understanding the generalizability of different design guidance.
Many cognitive or perceptual effects have been utilized by visualization research in recent years, such as the Dunning-Kruger effect~\cite{chen2024unmasking} and inattentional blindness~\cite{boger2021jurassic}.
Those effects can help understand the perceptual and cognitive limits of visual representations
and can improve the efficiency of visual data communication as a result~\cite{franconeri2021science, szafir2023visualization}.

For example, psychological \fix{phenomena} such as \textit{subitizing}---our ability to rapidly process a small (magic) number of objects~\cite{kaufman1949discrimination, miller1956magical}---could 
offer a theoretical basis for understanding why people may struggle with visualizations that involve 
many categories.
Subitizing is typically effective for quantities of up to four or five items. Beyond this range, individuals rely on counting, which is slower and more error-prone~\cite{piazza2002subitizing}.
Further, the relationship between grouping cues and subitizing, alternatively known as groupitizing~\cite{anobile2020groupitizing}, indicates that subitizing may play a role in categorical visualization tasks, where people draw conclusions about sets of marks rather than individual marks. 
The effectiveness of subitizing for less than six discrete items (or sets of items) may, in part, help account for the recommendations around encoding six categories in visualization design guidelines \cite{adobedesign}. 
Past studies have speculated that subitizing may play a role in visualization interpretation such as in visual search~\cite{haroz2012capacity}, reading pictographs~\cite{haroz2015isotype}, and estimating means~\cite{tseng2023evaluating}.
When people engage in subitizing versus counting or alternative comparative perceptual mechanisms may have
significant implications for how we choose to construct categorical visualizations. 
If we are biologically adapted to reason over six or fewer categories, we may need different design approaches for datasets with smaller category numbers than for those with larger category numbers. 

\begin{wrapfigure}{l}{0.25\textwidth}
    \centering
    \vspace{-0.5em}
    \includegraphics[width=\linewidth]{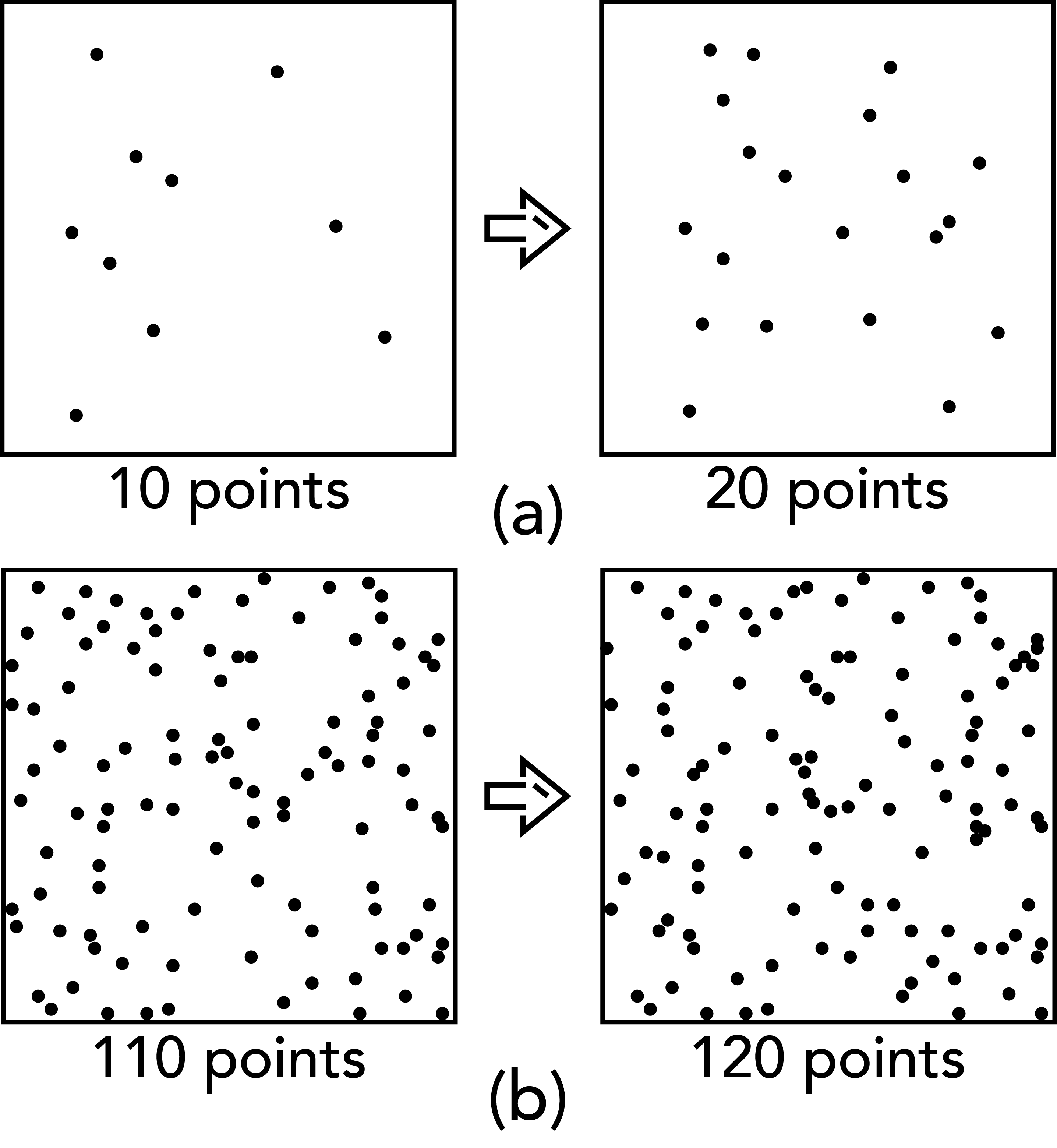}
    \vspace{-1.5em}
    \caption{Illustrating Fechner's Law with scatterplots. Charts in (a) and (b) have the same visual intensity (point number) difference, but the perceived difference in (a) is much stronger.}
    \label{fig:f-law}
    \vspace{-1em}
\end{wrapfigure}

While subitizing offers a hypothetical threshold for understanding design, Fechner's Law~\cite{fechner1948elements,harrison2014ranking,kay2015beyond} may help further explain performance patterns in larger numbers of categories.
The law posits that the perceived intensity of a stimulus grows logarithmically as its physical magnitude increases.
In the context of data visualization, this means that as the number of categories in a visualization increases, the perceptual difference between categories diminishes, making it harder for users to distinguish between them.
\add{\autoref{fig:f-law} illustrates the logarithmic differences of human perception in visualization using Fechner's Law.
The difference in the number of dots in the scatterplots 
is the same in each pair, from 10 to 20 in (a) and from 110 to 120 in (b).
However, the 
perceived difference in (a) is significantly higher than in (b).
}
This principle can explain why visualizations with a large number of categories often result in decreased accuracy and increased cognitive load, but that this decrease tends to become more gradual as the number of categories increases, even if the elements of a palette remain perceptually discriminable as in many design strategies for integrating larger numbers of colors into a palette \cite{graze2024building} or in randomly sampled palettes \cite{tseng2024revisiting}.

Our study builds on these perceptual and cognitive foundations by empirically testing how well people interpret categorical data visualizations, providing new insights into how to more effectively visualize categorical data through the theoretical lens offered by these mechanisms. The complexity of data visualizations compared to conventional perceptual studies prevents us from drawing firm causal connections between perceptual mechanisms and visualization design. We instead examine whether these mechanisms help us better model performance under different design conditions to ground actionable design guidance, understand how existing heuristics may generalize (or fail to generalize) to a range of applications, and offer translational research hypotheses for future work at the intersection of visualization and vision science.

\section{Experimental Overview}
\label{sec-methodology}

We conducted three experiments to understand how well people perform several tasks when visualizing data with different numbers of categories. Our objective is to capture patterns of performance across varying numbers of categories and use these patterns to better understand effective design practices through the lens of perceptual models.

\subsection{Task Selection}

We assess the effect of category number on visualization task performance using scatterplots as they are one of the most representative visualization types~\cite{quadri2021survey, munzner2014visualization, rensink2018information, gadhave2021predicting} and well-studied for categorical data~\cite{tseng2023evaluating, gleicher2013perception, sarikaya2018scatterplots}.
Scatterplot visualization tasks can be categorized into three levels: object-centric, browsing, and aggregate~\cite{sarikaya2018scatterplots}.
Since object-centric tasks focus on searching for specific objects rather than reasoning over sets of objects, we 
focused on browsing and aggregate tasks. 
We selected three representative tasks---a browsing-based class estimation task, an aggregate-based correlation comparison task, and an aggregate-based clustering task---
to assess the impact of categories across task complexities (see \autoref{fig:teaser} (b)):

\vspace{3pt}
\noindent \textbf{Class estimation} requires people to estimate the total number of classes that appeared in a scatterplot.
People must browse the whole plot to identify and count the number of classes~\cite{sarikaya2018scatterplots}.
\add{This class estimation task} correlates to the long-lasting number estimation task for evaluating subitizing in psychology~\cite{revkin2008does}, providing a task that may heavily leverage subitizing. 
Tasks correlated with those from cognitive science communities, 
like visual search~\cite{haroz2012capacity, gramazio2014relation} or centroid estimation \cite{hong2021weighted}, help provide translational insight into how people work with visualized data by facilitating connections between the two bodies of literature.

\vspace{3pt}
\noindent \textbf{Correlation comparison} requires people to choose the class with the highest correlation.
Comparing correlations is a traditional task in data visualization that has been extensively studied in single-class scatterplots~\cite{kay2015beyond, harrison2014ranking, rensink2010perception}.
This task requires people to aggregate data to identify correlations and make comparisons between classes~\cite{sarikaya2018scatterplots}.
However, increasing the number of categories encoded in the scatterplot
reduces people's 
correlation judgment accuracy
\cite{tseng2024shape}.
As hypothesized in past work, we anticipate that this decrease may be in part due to 
a reliance on subitizing in conjunction with other visual mechanisms to perform correlation comparisons.

\vspace{3pt}
\noindent \textbf{Clustering judgment} requires people to find a class that is the most tightly clustered among all classes.
Clustering judgment requires people to visually aggregate classes across the distribution~\cite{sarikaya2018scatterplots} and then compare the resulting classes. Clustering has significant implications for designing visual quality measures~\cite{jeon2023clams, wang2019improving, abbas2019clustme}.
However, people's comparison accuracy for clumpiness (higher clumpiness means more tightly clustered~\cite{wilkinson2005graph}) is significantly reduced as the number of categories increases~\cite{wang2019improving}, again correlating with the potential use of subitizing in selecting and characterizing individual classes.

\subsection{Overall Hypotheses}
Based on the three task designs and insights from prior research, we 
hypothesize:

\noindent \textbf{H1:} \textbf{Accuracy will remain stable for fewer than six categories.}

The first hypothesis 
pertains directly to how subitizing
impacts 
people's abilities to work with certain numbers of categories~\cite{revkin2008does}.
Previous studies have reported preliminary insights on unusual performance drops at six categories in visual search~\cite{haroz2012capacity} and mean judgment~\cite{tseng2023evaluating} tasks. Subitizing 
supports rapid, stable perception for four to six objects or sets of objects \cite{kaufman1949discrimination}.
Therefore, we anticipate the limits of subitizing may be within this range for visualizations, 
meaning people can accurately perceive and interpret categorical visualizations for six or fewer categories and that performance will decline sharply when the number of categories reaches six as people may rely on additional perceptual mechanisms to track and enumerate larger numbers of categories.

\noindent \textbf{H2:} \textbf{Accuracy will drop significantly for more than six categories.}

Fechner's Law states that the quality of a percept changes logarithmically with an increase in cognitive load~\cite{fechner1948elements}. 
We anticipate the cognitive load required to interpret categorical visualizations increases disproportionately as the number of categories exceeds six (the anticipated limit of subitizing), leading to significant differences in distribution for accuracy and efficiency with more than six categories.
More specifically, for a one-category change, Fechner's Law suggests that the decrease in performance may be more severe for category numbers near six (e.g., seven and eight) than for an equivalent step on higher numbers of categories (e.g., 12 and 13).

\noindent \textbf{H3:} \textbf{The impact of the number of categories on perceptual accuracy varies depending on the specific visualization task.}

Different tasks may require different cognitive processes~\cite{van2011effects} and analysis targets~\cite{sarikaya2018scatterplots}.
We anticipate tasks that require aggregation, such as clustering, may be more robust to increasing the number of categories than tasks that require browsing, like class estimation~\cite{chandler1991cognitive}. These aggregation-based tasks are more complex and likely rely on a larger set of perceptual mechanisms to achieve~\cite{albers2014task,szafir2016four}, meaning that the ability to leverage subitizing to select for and reason over classes plays a smaller role in overall performance.

\noindent \textbf{H4:} \textbf{Different encoding types may influence task performance as the number of categories increases.}

Shape has 
proven a more complex and hard-to-design visual encoding channel compared to color in categorical perception~\cite{tseng2024shape}. These differences influence people's abilities to distinguish categories and influence overall performance as a result. 
Similarly, we anticipate performance will be more robust on color-coded scatterplots than on shape-coded ones.

\noindent \textbf{H5:} \textbf{Increasing visual complexity in categorical visualizations exacerbates the decline in perceptual accuracy as the number of categories exceeds six.}

More complexity in visualizations leads to poor perception performance, such as when points become overdrawn~\cite{mayorga2013splatterplots}.
Increasing visualization complexity generally degrades overall performance. We anticipate an interaction between other aspects of visual complexity and the number of categories, 
with increases in both factors making it harder for people to accurately perceive and interpret data.

\section{Experimental Design}
We designed three experiments to test our hypotheses with each testing a different task. As these experiments leverage a number of common elements, we describe the experiments here and discuss their respective results in \autoref{sec-results}. 
The studies have been approved by [Redacted] Institutional Review Board.

\subsection{Experiment One: Class Estimation}
This experiment asked participants to estimate the total number of classes present in a series of scatterplots,
similar to methods employed in psychology experiments studying subitizing~\cite{revkin2008does}. Scatterplots varied in their encoding type (shape or color, between-participants), response duration (3 or 10 seconds, between-participants), complexity (low, medium, or high; within-participants), and number of categories ($N=2--15$, within-participants). Subitizing directly assists numerosity estimation, which is the core component of this task. Therefore, we anticipate that performance in this task will be well-modeled by subitizing limits, with people reliably able to identify six or fewer categories and performance degrading significantly with more than six categories.

\fix{While people might eventually count categories accurately with unlimited time, visualizations are designed to minimize cognitive load and facilitate rapid comprehension across a range of tasks~\cite{munzner2014visualization, szafir2023visualization}.
Many foundational tasks in visualization rely on a gist developed very quickly, followed by a more detailed analysis of target tasks, which leads to the importance of \emph{the speed of perception}~\cite{ryan1956speed}, more specifically, class estimation is one of those gist tasks: people get a rough sense of the number of classes and then dig into the classes sequentially for comparison, characterization, and other analyses.
}
While both subitizing and counting mechanisms have the potential to be highly accurate, \fix{counting takes significantly longer time~\cite{piazza2002subitizing}} and subitizing operates significantly faster~\cite{schleifer2011subitizing}. 
As a result, typical vision science studies 
investigate psychological phenomena like subitizing 
using very short time limits \fix{ranging from 500 ms to over 1500 ms~\cite{revkin2008does, vuokko2013cortical}}.
However, visualizations tend to operate over longer time scales and 
crowdsourcing platforms face
potential challenges in deploying time-constrained studies, such as differences in data loading and server connection times. 
To account for these differences, we 
measure performance over two different time limits: 
three seconds and 
ten seconds.
\fix{We implement the time limits as a between-subjects variable to reduce potential priming and learning effects (e.g., avoid people confusing
a three-second trial for a ten-second one).}
\add{These time limits were determined in piloting. }
We anticipate if subitizing 
is used to support multiclass scatterplot analysis, performance on small categories should be similar for both time limits, but will degrade with shorter exposure times for 
larger numbers of categories.

\subsubsection{Stimuli}
Scatterplots were rendered using a 400$\times$400 pixel graph.
The datasets contained between $N = 2--15$ categories.
We encoded categories using two of the most widely-used visual channels for representing data in multiclass scatterplots: color and shape.
To reduce the complexity of the study design, we 
used high-performing palettes from previous studies~\cite{tseng2023evaluating, tseng2024shape}, choosing the STATA S2 color palette and R shape palette to encode categories.
We followed other visual representation parameters from these studies, using a filled mark with a three-pixel radius for color-encoded plots and a 6$\times$6-pixel window for shape encodings. Colors and shapes were uniquely assigned to each category using random selection.

\begin{figure*}[t]
    \centering
    \includegraphics[width=0.75\linewidth]{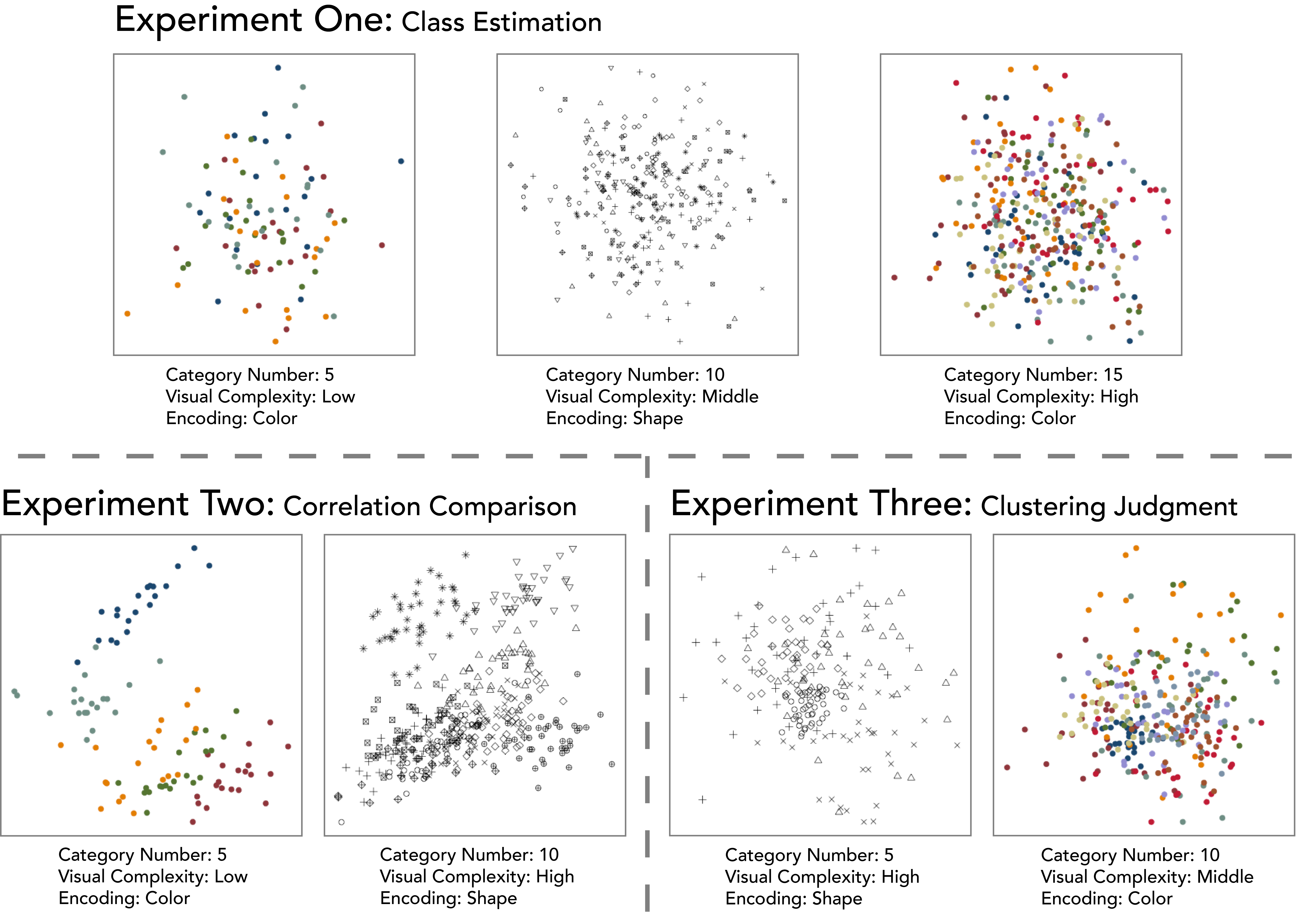}
    \caption{Example stimuli of scatterplots applied in class estimation task (Experiment One), correlation comparison task (Experiment Two), and clustering judgment task (Experiment Three), with varying visual encodings, number of categories, and visual complexities.}
    \label{fig:stimuli1}
    \label{fig:stimuli2}
    \label{fig:stimuli3}
    \vspace{-1em}
\end{figure*}

Class data in the employed datasets was drawn from 2D Gaussian normal distributions with variances ranging from (0, 1), with random mean x- and y-values 
between [0.1, 0.9].
Further, we created three different visual complexity levels for the classes by controlling the number of points from each class: low (20 points), middle (30 points), and high (40 points).
Each scatterplot only contains classes with all low, middle, or high numbers of points.
See \autoref{fig:stimuli1} for examples of scatterplot stimuli used in Experiment One.

\subsubsection{Procedure}
Our experiment consisted of three or four phases, depending on the design condition: (1) informed consent, (2) color vision deficiency (CVD) screening using Ishihara plates for participants viewing color-coded scatterplots, (3) task description and tutorial, and (4) formal study.
Each participant 
only saw either all color-coded scatterplots or all shape-coded scatterplots and were assigned to either the 3 or 10-second condition for all trials.

Participants first provided informed consent under our IRB protocol and then provided basic demographics.
Participants who saw color-coded scatterplots 
then needed to successfully pass the CVD assessment.
Afterward, participants were introduced to the class estimation task description and led to the tutorial section. They completed three tutorial questions, asking them to complete the target task with 2 to 3 categories with color or shape encodings, whichever 
aligned with their assigned condition.
They were required to successfully answer all  tutorial questions before proceeding to reduce possible ambiguities in task understanding. 

During the formal study, participants completed our target task (``estimate the total number of categories in this scatterplot'') for 42 stimuli presented sequentially in a random order.
The 42 stimuli included one scatterplot for each combination of 14 different category numbers (2-15) and three visual complexity levels (high, middle, low).
We added three simple two-class scatterplots, showing two well-separated compact classes,
as engagement checks following other studies~\cite{tseng2023evaluating}.
Participants had either three seconds or ten seconds to view each stimulus, depending on their assigned condition. Then the stimulus was hidden, and participants were required to provide an answer \fix{by typing in a number in an input textbox}.

\subsubsection{Participants}
We recruited 239 participants on Amazon Mechanical Turk (MTurk) with at least a 95\% approval rating and located within the US and Canada. 
All participants reported normal or corrected-to-normal vision.
34 participants who failed more than one engagement check were excluded, resulting in an 86\% acceptance rate.
Among the remaining 205 users (133 male, 72 female; 22–60 years of age), 53 participants saw color-coded scatterplots for three seconds, 54 saw color-coded scatterplots for ten seconds, 
50 saw shape-coded scatterplots for three seconds, and 48 
saw shape-coded for ten seconds. 
These studies took 4--7 minutes on average.

\subsection{Experiment Two: Correlation Comparison}
This experiment asked participants to pick
the class that had the highest correlation. 
This task requires people to both select between different classes and then aggregate properties across those classes, potentially involving a larger number of perceptual mechanisms than Experiment One. Scatterplots varied in their encoding type (shape or color, between-participants), complexity (low, medium, or high; within-participants), and number of categories ($N=2--15$, within-participants).
\add{This study partially replicates past studies that demonstrated a potential subitizing effect \cite{tseng2023evaluating,tseng2024shape}
to determine whether the prior effects replicate and, if so, to evaluate the potential connection to subitizing.
}

\subsubsection{Stimuli}
The scatterplot stimulus design and complexity implementation was the same as in Experiment One; however, our data generation manipulated correlation to control task difficulty. 
Following the process outlined in Tseng et al.~\cite{tseng2024shape}, we employed Pearson's correlation coefficient to control per-category correlations.
The scatterplots were generated by the random multivariate normal data generation function in R~\cite{ripley2013package}, sampling random x- and y-mean values ranging between [0.1, 0.9].

In prior correlation studies for single-class scatterplots, the just-noticeable difference (JND) of correlations was reported to range from 0.05 to 0.15~\cite{kay2015beyond, harrison2014ranking, rensink2010perception}.
Therefore, we set the category with the highest correlation to have a [0.8, 0.9] covariance and the second-highest category has at least a 0.1 lower correlation difference.
We jittered points to avoid overlap and resampled points until the correlation coefficient values satisfied these criteria.
See \autoref{fig:stimuli2} for stimuli of scatterplots used in Experiment Two.

\subsubsection{Procedure}
Participants followed the same general procedure as in Experiment One.
Each participant completed 42 formal trials (one for each combination of category number and complexity level) and 3 engagement checks. Participants responded to the target task (``identify which category is the most correlated'') by choosing the target color or shape from a set of radio buttons 
including all colors or shapes shown in the plot~\cite{tseng2023evaluating}. Participants were given up to 30 seconds to respond to each stimulus (duration determined in piloting).

\subsubsection{Participants}
We recruited 112 participants on MTurk with at least a 95\% approval rating and located within the US and Canada.
All participants reported normal or corrected to normal vision.
14 participants who failed more than one engagement check were excluded, resulting in an 87.5\% acceptance rate.
Among the remaining 98 participants (59 male, 39 female; 22–60 years of age), 49 saw color-coded scatterplots and 49 saw shape-coded scatterplots.
This study took 12 minutes on average.

\subsection{Experiment Three: Clustering Judgment}
The third experiment asked people to choose the most tightly clustered class. 
People again identified different categories and drew inferences about the general shape formed by the boundaries of the class. Scatterplots varied in their encoding type (shape or color, between-participants), complexity (low, medium, or high; within-participants), and number of categories ($N=2-15$, within-participants).

\subsubsection{Stimuli}
The stimuli again were generated using the same general parameters as in Experiment One. 
Cluster tightness was controlled using x- and y-variance.
The most tightly clustered and second most tightly clustered classes differed in this variance by 0.1 along each dimension to allow perceptual differences between clusters~\cite{wang2019improving}, with the most tightly clustered ranging in x- and y-variance from [0.1, 0.2].
Category points were sampled using a 2D Gaussian distribution. 
As in previous experiments, the x- and y-mean values were randomly sampled between [0.1, 0.9].
See \autoref{fig:stimuli3} for examples of scatterplots used in Experiment Three.

\subsubsection{Procedure}
We employed the same general procedure as Experiments One and Two.
Each participant completed the target task (``identify which category is the most clustered'') for 42 formal trials and 3 engagement checks in a random sequential order.
Participants selected the target category as a radio button 
with the corresponding shape or color. Participants had 30 seconds to respond to each stimuli (duration determined in piloting). 

\subsubsection{Participants}
We recruited 112 participants on MTurk with at least a 95\% approval rating and located within the US and Canada.
All participants reported normal or corrected to normal vision.
10 participants who failed more than one engagement check were excluded, resulting in a 91\% acceptance rate.
Among the remaining 102 participants (70 male, 32 female; 24–64 years of age), 53 were in the color-coded group, and 49 were in the shape-coded group.
This study took 12 minutes on average.

\section{Results}
\label{sec-results}

We discuss significant results and statistical analysis 
using both traditional inferential measures and 95\% bootstrapped confidence intervals ($\pm$ 95\% CI) for fair statistical communication~\cite{dragicevic2016fair}.
\fix{We used accuracy as our dependent measure, aggregated across all participants within each experiment as it's a binary response (correct or not) for each trial.}
We computed \fix{three-way ANOVAs for each experimental task individually} to compare the impact of three different independent factors on accuracy: category number, encoding type, and visual complexity (i.e., the number of points).
The data were approximately normally distributed in the results of each experimental setting.
We found no significant two-way interactions
and, as a result, do not discuss these here.
Please see the \href{https://osf.io/y3z2b/?view_only=7f6569187b344fadbd11cc09a6e63d24}{OSF supplements} for data and results \fix{including full ANOVA tables}.
\autoref{tab:anova1} \fix{summarizes 
the results for all our experiments---\emph{T1-3s}, \emph{T1-10s}, \emph{T2}, and \emph{T3}}.
 
\begin{table}[ht]
\centering
\caption{\fix{Main effects of the ANOVA 
for each of our experiments with three factors: category number (\textit{CN}), encoding type (\textit{Encoding}), and visual complexity (\textit{VC}).}
Significant effects are shown in \textbf{bold}.}
\label{tab:anova1}
\begin{tabular}{lrrrrr}
  \hline
\textbf{\emph{T1-3s}}  & Df & Sum Sq & Mean Sq & F value & Pr($>$F) \\ 
  \hline
CN & 1 & 8.34 & 8.34 & 627.37 & \textbf{<.0001} \\ 
  Encoding & 1 & 0.05 & 0.05 & 3.58 & 0.0624 \\ 
  VC & 2 & 0.24 & 0.12 & 9.00 & \textbf{0.0003} \\ 
  \hline
  Residuals & 72 & 0.96 & 0.01 &  &  \\ 
  \hline
  \hline
\textbf{\emph{T1-10s}}  & Df & Sum Sq & Mean Sq & F value & Pr($>$F) \\ 
  \hline
CN & 1 & 7.98 & 7.98 & 593.81 & \textbf{<.0001} \\ 
  Encoding & 1 & 0.02 & 0.02 & 1.32 & 0.2546 \\ 
  VC & 2 & 0.53 & 0.26 & 19.64 & \textbf{<.0001} \\ 
  \hline
  Residuals & 72 & 0.97 & 0.01 &  &  \\ 
   \hline
   \hline
\textbf{\emph{T2}} & Df & Sum Sq & Mean Sq & F value & Pr($>$F) \\ 
  \hline
CN & 1 & 2.56 & 2.56 & 325.12 & \textbf{<.0001} \\ 
  Encoding & 1 & 0.29 & 0.29 & 36.93 & \textbf{<.0001} \\ 
  VC & 2 & 0.04 & 0.02 & 2.57 & 0.0834 \\ 
  \hline
  Residuals & 72 & 0.57 & 0.01 &  &  \\ 
   \hline
   \hline
\textbf{\emph{T3}}  & Df & Sum Sq & Mean Sq & F value & Pr($>$F) \\ 
  \hline
CN & 1 & 3.32 & 3.32 & 487.00 & \textbf{<.0001} \\ 
  Encoding & 1 & 0.35 & 0.35 & 52.01 & \textbf{<.0001} \\ 
  VC & 2 & 0.03 & 0.02 & 2.33 & 0.1046 \\ 
  \hline
  Residuals & 72 & 0.49 & 0.01 &  &  \\ 
   \hline
\end{tabular}
\vspace{-0.5em}
\end{table}

\subsection{Subitizing Limit and 
Category Number}
\label{sec-analysis-cat}

Our results support \textbf{H1} and partly support \textbf{H2}:
for all tasks, we found that people's accuracy was relatively stable at fewer than six categories and dropped significantly after six categories (H1), but performance 
reduced when the category numbers became very high (H2).

\begin{figure*}[htbp]
    \centering
    \includegraphics[width=\linewidth]{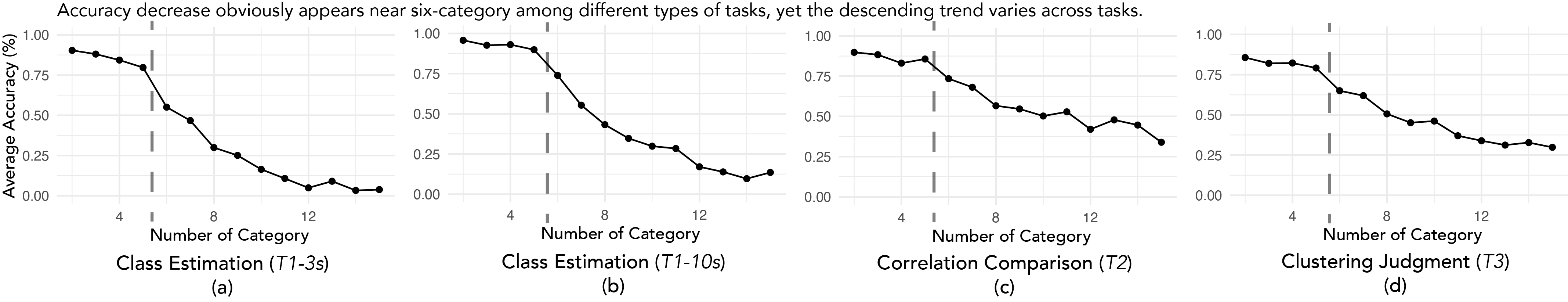}
    \vspace{-1.5em}
    \caption{
    Overall accuracy distributions in three experiments, separated by four different task settings.
    The x-axis shows the number of categories in the scatterplots, and the y-axis is the average accuracy in each task.
    The dashed gray lines denote the separation between five and six categories.}
    \vspace{-1em}
    \label{fig:accexp1}
\end{figure*}

Increasing category numbers significantly reduced accuracy (\fix{($p < .0001$ for all settings}). 
To further validate subitizing's impact across different tasks, \autoref{fig:accexp1} shows accuracy changes across category numbers on the three experiments with four task settings (3s for class estimation, 10s for class estimation, correlation comparison, and clustering).
Performance drops between five and six categories for all four task settings, as denoted by the gray dashed lines.

We found 
a significant difference in average accuracy between five to six categories among all tasks:
24\% for estimation at 3s, 16\% for 10s estimation, 12\% for correlation, and 14\% for clustering.
This observation indicates five to six categories could correspond to potential subitizing limits in visualizations, aligning with insights from cognitive science~\cite{piazza2002subitizing}.
Therefore, we further explored the 
data patterns on two category ranges: $N < 6$ and $N \ge 6$.

For fewer than six categories, 
performance drops were small for
increasing category numbers, at only 3-4\% for each added category across 
all three experiments.
Average accuracy remained 
relatively high.
For example, even with restricted three-second viewing time limits, people still achieved nearly 80\% accuracy in judging the number of classes in scatterplots with five categories (\autoref{fig:accexp1} (a)).
This insight aligns with previous research where people achieved robust performance (nearly 90\%) at five categories in mean judgment tasks~\cite{tseng2023evaluating}.

However, for six or more categories, 
the performance 
distribution significantly changed.
We observed the 
strongest performance drops when increasing %
from five to eight categories, showing an average accuracy reduction per added category at around 16\% for \autoref{fig:accexp1} class estimation tasks and around 8\% for correlation and clustering.
However, as category numbers continued to increase, 
the effects of adding more categories 
on accuracy were reduced despite class encodings remaining discriminable.
This effect was especially notable for 
category numbers greater
than 12, where performance appears to essentially plateau.
The observed two-segment performance descending distribution (decreasing sharply after six and slowing to near stable performance after 12)
could be at least in part modeled by Fechner's Law.
However, we cannot find a single inflection point across all tasks that can separate the fast-descending and flat-descending ranges of category numbers to indicate a consistent logarithmic model.
This finding indicates the strength of the hypothesized effects may be different across tasks: in our case, it is clearer in browsing-based tasks (which more closely resemble subitizing tasks in psychology) than in aggregate-based tasks (which likely involve a larger set of mechanisms \cite{szafir2016four}).

Overall, these results show that the distribution of accuracy changes notably for more than six categories, lending empirical support to traditional design heuristics around the limits of categorical encodings. 
Before reaching this limit, accuracy remains relatively stable and high, indicating that the perceptual system can manage up to six categories effectively. 
Beyond this point, the decrease in accuracy reduces at a rate correlated with Fechner's Law. However, the rate of decline varies across tasks, 
suggesting that other mechanisms may be at play for different tasks and the performance of these mechanisms degrades at different rates than those predicted by subitizing.

\subsection{Higher Task Complexity is More Robust to Increasing Categories}
\label{sec-analysis-tc}

Our results support \textbf{H3}:
we found the impact of category numbers on perceptual accuracy varied depending on the specific visualization task.

\fix{To explore the impact of task settings}, we conducted simple slope comparisons with Bonferroni correction as a post-hoc analysis.
In addition to the general results shown in \autoref{fig:accexp1}, 
there was a significant difference between \textit{T1-10s} and \textit{T2} ($t=-8.515$, $p<.0001$), 
\textit{T1-10s} and \textit{T3} ($t=-6.961$, $p<.0001$), 
\textit{T1-3s} and \textit{T2} ($t=-8.949$, $p<.0001$), and 
\textit{T1-3s} and \textit{T3} ($t=-7.396$, $p<.0001$).
People's judgment accuracy on \textit{T1-3s} or \textit{T1-10s} have a significantly steeper slope than both \textit{T2} and \textit{T3}.
In other words, the same increase in category numbers will result in 
stronger performance drops in 
the class estimation task than both correlation comparison and clustering tasks.

This pattern 
suggests that aggregation (T2 and T3) may actually be more robust to increases in the number of categories compared to browsing tasks like class estimation.
This may also align with cognitive insights that visual aggregation usually relies more upon complex cognitive processes like ensemble coding~\cite{ratwani2008thinking, albers2014task,szafir2016four}.
\fix{
We note that the 
data distributions differed across experiments to 
control task difficulty, creating a potential
confounding effect between task and stimulus design.
While our focus on relative performance within each task mitigates the potential impact on our findings, future work should consider examining different tasks using the same data distributions.}

\begin{figure*}[htbp]
    \centering
    \includegraphics[width=\linewidth]{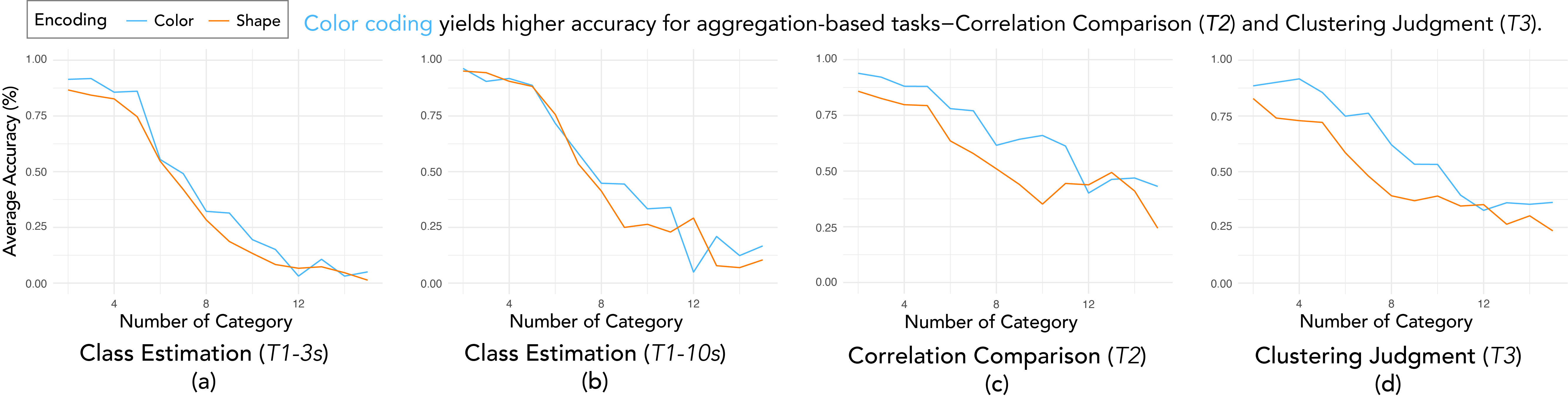}
    \vspace{-1.5em}
    \caption{Accuracy comparisons with color and shape encodings on four different task settings. (a) and (b): class estimation tasks with 3s and 10s time limits. (c): correlation comparison task. (d): clustering judgment task.}
    \label{fig:encodingcomp}
    \vspace{-1em}
\end{figure*}

\subsection{Color is 
More Effective for Aggregation Tasks}

Our results partly support \textbf{H4}:
color encodings 
were more robust than shape encodings to increasing numbers of categories in aggregate-based tasks (clustering and correlation comparison).
We found a significant effect of category encoding types on accuracy \fix{in tasks \textit{T2} ($F(1,71) = 36.93, p < .0001$) and \textit{T3} ($F(1,71) = 52.01, p < .0001$), but did find any significant effect in tasks \textit{T1-3s} and \textit{T1-10s}.}

\autoref{fig:encodingcomp} illustrates the accuracy results and \autoref{fig:encoding} (a) shows the post-hoc analysis results of two category encodings with specific task types.
These results show that the accuracy of color encodings is generally higher than shape encodings among all tasks.
However, we also observed that for browsing-based class estimation tasks (see \autoref{fig:encodingcomp} (a) and (b)), the accuracy distributions of color and shape encodings cannot be distinguished clearly from each other. 
For aggregation-based correlation comparison and clustering tasks (see \autoref{fig:encodingcomp} (c) and (d)), the accuracy differences between color and shape encodings are significant at $t=6.077$, $p<.0001$ (\textit{T2}) and $t=7.212$, $p<.0001$ (\textit{T3}) respectively. 
This suggests that color could be a more perceptually effective encoding than shape when rendering categorical data, but only for certain kinds of tasks. 

\subsection{Reducing Point Numbers May Not Significantly Improve Performance}

Our results do not support \textbf{H5}:
lower visual complexity failed to significantly improve 
accuracy when modeled as
varying point number.
We found a significant effect of visual complexity on accuracy \fix{in tasks \textit{T1-3s} ($F(1,71) = 9.00, p = .0003$) and \textit{T1-10s} ($F(1,71) = 19.64, p < .0001$);}
 however, that effect did not match our hypothesized outcomes.

\begin{figure}[htbp]
    \centering
    \includegraphics[width=0.95\linewidth]{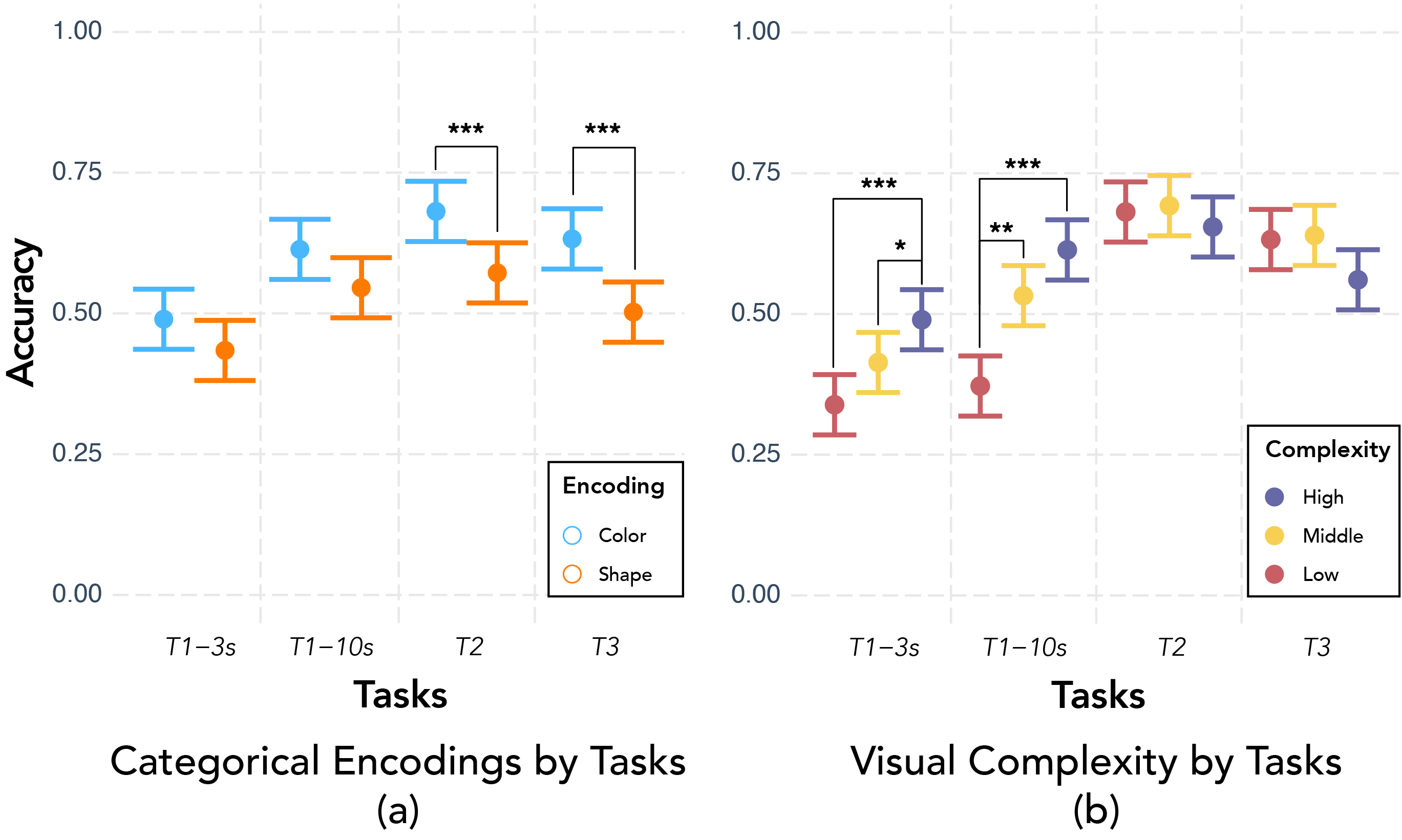}
    \vspace{-0.5em}
    \caption{(a) 
    shows the \fix{post-hoc results comparison with 95\% confidence intervals (CI)} per encodings with tasks.
    (b) 
    plots the \fix{post-hoc results comparison with 95\% CI} per visual complexities with tasks.
    \fix{Significant differences are denoted as \textbf{*} ($p<.05$), \textbf{**} ($p<.01$), and \textbf{***} ($p<.0001$).}
    }

    \label{fig:encoding}
    \label{fig:complex}
    \vspace{-1em}
\end{figure}

\autoref{fig:complex} (b) shows the post-hoc analysis results between complexity and task. 
The general pattern for aggregation-based tasks (\textit{T2} and \textit{T3} in \autoref{fig:complex} (b)) aligns with our assumption that 
accuracy for lower visual complexities is slightly higher than higher complexities; however, the differences are not significant.
For browsing-based class estimation tasks (\textit{T1-3s} and \textit{T1-10s} in \autoref{fig:complex} (b)),  
low complexity led to lower accuracy compared to middle and high complexity.

Neither of these results 
support the 
hypothesis that reducing visual complexity leads to improved perceptual accuracy.
This effect was especially notable for browsing-based tasks, where the low points number may cause their overall distributions to be too sparse for people to make immediate judgments.
The limited correlation between complexity and performance matches observations from Tseng et al. \cite{tseng2023evaluating}.
Complexity factors beyond point number, such as patterns in distribution density, may lead to different outcomes.
Future studies should better explore this impact using visual quality measures that align with human perception~\cite{wang2019improving, jeon2023clams, abbas2019clustme} to guide visual complexity control.

\section{Discussion}
\label{sec-discussion}

We explored how varying category numbers influence accuracy on three multiclass scatterplot tasks, with an emphasis on how two relevant psychological phenomena and theory---subitizing and Fechner's Law---may explain patterns in categorical data interpretation. 
Our results provide further evidence of the role of subitizing in data visualization, offer new perspectives reflecting on prior heuristics and research findings, and suggest actionable design guidelines and future research opportunities.

\subsection{Viewing Data Visualization via the Lens of Psychology}

Our results identify
a potential role of subitizing in visualization---supporting categorical reasoning---to
help predict how 
category number affects people's abilities to use visualizations effectively. 
Prior research in cognitive psychology has 
established that subitizing is effective for small numbers of items, beyond which individuals must rely on other, often less efficient, cognitive processes such as counting~\cite{kaufman1949discrimination, piazza2002subitizing}.
Our study extends these findings by demonstrating that these limits directly affect  task accuracy in categorical visualizations.
More specifically, our results confirm that when the number of categories reaches six, people start to experience significant performance drops for a range of categorical tasks, providing empirical and theoretical support for past design heuristics and highlighting the importance of considering these cognitive boundaries for visualizations.

This result aligns with findings from related studies that found sudden performance drops 
for tasks such as mean judgments~\cite{tseng2023evaluating,tseng2024revisiting}, visual search~\cite{haroz2012capacity}, isotype visualizations~\cite{haroz2015isotype}, and correlation judgments~\cite{tseng2024shape}. 
When viewed as an outcome of subitizing, such performance drops 
imply a more general visualization phenomenon 
described by known constraints of subitizing for $N < 6$ and by Fechner's Law for $N \ge 6$. That is, people are able to quickly and efficiently identify and reason across six or fewer encoded categories (\textit{subitizing}) and, after a certain threshold (approximately 12 
in our results), 
additional categories have little to no effect on analysis across categories (Fechner's Law).

\fix{These results, as a whole, can further 
connect to 
\emph{Dual-System Theory} in psychology~\cite{frankish2010dual}, which 
suggest individuals have two different sets of decision-making processes, System 1---impulsive, fast, and acts without thinking---and System 2---a more cognitive, deliberate, thinking process.
Subitizing, the rapid and accurate perception of small quantities, aligns with System 1 processing,
characterized by its automatic and effortless nature.
When quantities exceed the subitizing limit, processing shifts to the more effortful and conscious System 2, involving deliberate counting or estimation.}

Our results also help explain tensions in prior studies. Tseng et al.~\cite{tseng2023evaluating} 
failed to confirm the guideline proposed by Gleicher et al.~\cite{gleicher2013perception}, which reported adding additional distractor classes does not significantly impact mean judgment performance in multiclass scatterplots.
Considering visualization through the lens of subitizing explains this contradiction:
Gleicher et al.'s study
only tests category numbers within the subitizing range ($N \le 3$), so their results confirmed the subitizing ability of near-instant categorical perception. Alternatively, 
Tseng et al.~\cite{tseng2023evaluating} explored beyond the subitizing range ($N \le 10$)
and observed the performance decrease that may be explained by Fechner's Law under more complex cognitive processes such as counting or aggregation.

\subsection{\add{In Dialogue with} Existing Heuristic Design Guidelines}

Many 
design heuristics suggest limiting the number of categories to avoid overwhelming the viewer.
However, empirical evidence supporting these guidelines has been sparse, and both research and commercial tools offer strategies for pushing encodings beyond these limits \cite{graze2024building}. By evaluating these heuristics using psychological models, we can apply an empirical lens to better understand if and when this and other heuristics hold in practice.

Our study provides both empirical evidence and a theoretical basis for the limit of six.
We found that 
accuracy in 
categorical analysis tasks remained stable for smaller categories but declined 
sharply with 
more than six categories,
providing empirical support for the up-to-six claims from prior heuristics~\cite{adobedesign, graze2024building}.
The six-category baseline, when viewed in relation to subitizing, is likely to generalize well across most visualization tasks that require selecting and comparing features of different groups. Further, characterizing performance through subitizing and Fechner's Law enables designers to make more informed predictions about how effectiveness will change as they increase the number of categories they choose to encode. While encoding more than six categories leads to lower overall performance, people still are capable of analyzing data at these scales, and the performance degradation from adding more categories quickly tapers off. This fall-off implies that people can make sense of larger numbers of categories (if less efficiently) and that visualizations are unlikely to benefit from reducing category count unless the data can be reduced to six or fewer categories.

\add{
While the six category rule is among the most common design heuristics employing a prescriptive quantity for design, other heuristics also raise specific design thresholds that subitizing and Fechner's Law may in part explain. }
\add{
For example, in palette design, an up-to-10 suggestion states ``\emph{If you add too many colors to the palette, it’ll be difficult to comprehend the chart.}''~\cite{tenfusion}.}
\add{
Others note that it may be difficult to find more than eight distinctive colors for categorical palettes~\cite{wong2010points}. This guidance contradicts the existence of at least thirteen readily-namable, readily distinguished colors \cite{berlin1991basic}
as well as our ability to sample far more than eight distinguishable colors from a limited numerical color space 
\cite{szafir2018modeling}; however, this heuristic does align with subitizing limits. Reasoning across more than eight color-coded categories is likely more difficult due to an inability to efficiently leverage subitizing than to an inability to distinguish between the colors themselves.
}

\add{
Other heuristics for design beyond categorical data emphasize the potential negative impact of excess visual features in visualizations, such as how text color, 
gridlines, and backgrounds may interfere with chart perception~\cite{munzner2014visualization}.
In past studies from psychology, adding additional distracting elements \cite{goldfarb2013counting} or additional encodings \cite{trick2008more} may slow subitizing, meaning distracting visual information may also interfere with categorical analysis.
Certain encoding types may use visual features that are well-suited to the cognitive processes used to accomplish different tasks. 
Understanding the connections between features and task-relevant processes can help guide more effective task-driven design guidelines \cite{kim2018assessing,albers2014task,saket2018evaluating}. 
For example, we found that more complex aggregation tasks that likely leverage other processes in addition to subitizing performed better with color than with shape, which may imply colors are efficiently processed by these processes. As people may process these channels differently \cite{albers2014task}, future work should explore if the perceptual mechanisms we use to process particular visual features make them well-correlated to certain tasks.
}

\subsection{Design and Research Implications}

\fix{Our results provide recommendations to promote future visualization design and research in general. Given our stimuli, we 
primarily focus on visual data communication to inform 
static graphs.}

\vspace{3pt}
\noindent\textbf{When possible, limit visualizations to six categories:}
Aligned with several existing heuristic guidelines, this recommendation 
reflects likely subitizing limits in data visualization. 
Above six categories, people will be less accurate and slower in analyzing data as they are likely to need to engage less efficient perceptual mechanisms to reason over categories. 
When the number of categories surpasses this threshold, people experience a significant decline in perceptual accuracy, leading to more potential data misinterpretation.

\vspace{3pt}
\noindent\textbf{Higher category numbers are more robust with aggregation-based tasks:}
Our findings suggest that category number's impact on perceptual accuracy varies depending on specific tasks at hand when exceeding the subitizing limit.
For aggregate-based tasks like clustering, 
people may 
be able to process a higher number of categories with relatively high accuracy assuming a well-designed visual encoding. This is likely due, at least in part, to additional mechanisms that are not subject to the same limits as subitizing playing a larger role in these tasks.

\vspace{3pt}
\noindent\textbf{Prioritize color encodings:}
We
confirm past findings \cite{tseng2024shape} that color is more easily distinguishable than shape when encoding categorical data.
While color encodings may have accessibility limitations, if these limitations can be minimized or avoided, color provides a more robust and effective categorical encoding channel across a variety of tasks.

\vspace{3pt}
\noindent\textbf{Psychology can provide theoretical grounding for design heuristics:}
Integrating advances in psychology into data visualization research offers a promising interdisciplinary way to improve the field~\cite{elliott2020design, bae2025bridging}.
Our study further confirms the value of using these grounded theories as guidance to inform design guidelines, which aligns with previous visualization research that investigates the application of psychological theories in visualization design and interpretation, like Weber's Law~\cite{harrison2014ranking, soni2018perception} and confirmation bias~\cite{wang2024causal, li2025confirmation}.
These theories help \emph{ground, refine, and generalize heuristic guidance} by bridging design practice with the processes by which people interpret visualizations. Connecting visualization interpretation to perceptual processes offers new hypotheses for vision science, such as opportunities to understand how different mechanisms work together to influence different kinds of data interpretation (e.g., different tasks), therefore building up real empirical foundations~\cite{kosara2016empire}. 
However, our results stress the importance of translational empirical research that confirms the effects these theories have in practice within visualizations. For example, subitizing and Fechner's Law help model performance across tasks, but are insufficient to fully explain performance differences as a function of category number. 

\fix{Our work reveals that understanding the underlying psychological mechanisms behind visualization heuristics offers more than just empirical validation; it provides genuine explanatory and predictive value.
By studying these mechanisms, researchers can determine the boundaries, edge cases, and potential conflicts between heuristics, leading to more robust and generalizable visualization design guidelines.}
We hope our work can promote future interdisciplinary efforts to foster a better understanding of perception in the context of data visualization to advance both visualization and vision science research.

\subsection{Limitations \& Future Work}

Visualization types other than scatterplots, such as bar charts, pie charts, or heatmaps, may present different challenges and opportunities for graphical perception~\cite{quadri2024do} and may offer different insights into the role of various perceptual laws and functions. 
Further, we only investigated a limited set of categorical encodings. 
Even though these palettes were chosen based on results in previous studies, they do not encompass the full range of possible encoding options available to designers.
For instance, other palettes and encoding channels, such as textures~\cite{interrante2000harnessing}, size~\cite{smart2019measuring}, and position~\cite{hong2021weighted}, could also benefit perception efficiency.

Redundant encoding---using multiple visual channels to represent the same factors in data---can %
enhance accuracy for visualization tasks~\cite{nothelfer2017redundant, gleicher2013perception}.
Our study did not investigate the potential benefits of redundant encodings, such as using both color and shape simultaneously to encode one category~\cite{smart2019measuring}.
This approach may benefit task performance, particularly in scenarios involving higher category numbers, and may offer additional features that the visual system can reason over simultaneously.

While our study focused on three specific tasks, these represent only a subset of the tasks that people commonly perform when interacting with categorical visualizations~\cite{sarikaya2018scatterplots}.
These tasks specifically require distinguishing different classes of points and analyzing data across these classes. 
Other tasks, such as identifying outliers, comparing skewness, or performing time-series analysis, may place different demands on people's perceptual and cognitive resources and may require using class information in different ways.
Further, these tasks may help provide insight into whether there is a causal or correlative relationship between subitizing and multiclass analysis, as well as other mechanisms that might be at play. 
Future work should explore a wider range of tasks to better understand subitizing and the types of tasks it helps model.
\fix{We aimed to cover a large range of category numbers, which reduced the trials for participants at each number, resulting in a lack of insights into individual differences which should be studied in the future.}

\section{Conclusion}
\label{sec-conclusion}

This study deconstructs design heuristics and visualization perception with psychological phenomena.
More specifically, our study provides a better understanding of data visualization in relation to \emph{subitizing} \fix{phenomena}. 
By empirically investigating the effects of category number, visual encoding, and task complexity, we have identified critical thresholds and patterns in behavior that suggest the use of subitizing in interpreting categorical visualizations. Performance patterns indicate that people's abilities to make sense of multiclass data change at six categories, beyond which perceptual accuracy declines significantly and correlates with behaviors predicted by Fechner's Law.
Our results culminate in a set of empirically grounded design guidelines that can inform more effective categorical visualizations.
These design guidelines and implications offer a roadmap for creating more effective visualizations for categorical data.
By adhering to these principles, designers can ensure that their visualizations align with the cognitive capabilities of their users, thereby maximizing both accuracy and usability.
Our work also demonstrates the significance of connecting theories from cognitive psychology and practices in data visualization, offering practical guidelines that enhance the effectiveness of visualizations and insight into how and why guidelines may generalize.

\acknowledgments{
We thank the reviewers for their insightful comments.
This work was supported by NSF IIS \#2046725 and by NSF CNS \#2127309 to the Computing Research Association for the CI-Fellows Project.
}

\bibliographystyle{abbrv-doi-hyperref-narrow}

\bibliography{new}

\end{document}